\documentstyle[11pt,newpasp,twoside,epsf]{article}
\def\edcomment#1{\iffalse\marginpar{\raggedright\sl#1\/}\else\relax\fi}
\reversemarginpar

\begin{document}
\title{Where has all the polarization gone?}
 \author{E. Middelberg, A. L. Roy, U. Bach}
\affil{Max-Planck-Institut f\"ur Radioastronomie, Bonn, Germany}
\author{D. C. Gabuzda}
\affil{Physics Department, University College Cork, Ireland}
\author{T. Beckert}
\affil{Max-Planck-Institut f\"ur Radioastronomie}

\section{Introduction}

Circumnuclear tori are a central ingredient in the unification of the
AGN phenomenon, but the conditions in the tori, the jet collimation,
and the accretion mechanisms are still poorly constrained. Magnetic
fields are involved in jet collimation and probably in feeding
material into the nucleus, but those that are derived using
equipartition are uncertain since equipartition conditions are not
known to hold.

A more direct measurement of magnetic field strength can be made using
Faraday rotation (FR) and free-free absorption (FFA). FR changes the
electric vector position angle of a polarized wave passing through a
magnetized plasma by an angle $\chi = R_{\rm m}\,\lambda^2$, where the
rotation measure $R_{\rm m}$ is the path integral over the
line-of-sight component of the magnetic field, $B_{\parallel}$, and
the density of thermal electrons, $n_{\rm e}$. FFA depends on $n_{\rm
e}$, the path length, $L$, and the electron temperature, $T_{\rm
e}$. The value of $L$ can be estimated by assuming it to be the same
as the width of the region of FFA in the VLBI images. Making
reasonable assumptions about $T_{\rm e}$, estimates of $n_{\rm e}$ can
also be derived from the FFA measurements. Thus, a joint analysis of
FR and FFA measurements can provide direct diagnostics of the
magnetic-field strength $B_{\parallel}$ with minimum imposed
assumptions.

\section{Sample and Results}
We selected all five AGNs that we found in the literature that had
steeply rising spectra across parts of the jet at pc scales. In all
cases, the absorption was most likely FFA due to a pc-scale foreground
absorber, perhaps the ionized inner edge of an obscuring torus or an
accretion flow. The sample comprises NGC~1052 (LINER), NGC~4261
(FR\,I), Centaurus~A (FR\,I), Hydra~A (FR\,I) and Cygnus~A
(FR\,II). Polarimetric observations were carried out with the
VLBA\footnote{The National Radio Astronomy Observatory is a facility
of the National Science Foundation, operated under cooperative
agreement by Associated Universities, Inc.} at 15.4\,GHz with 60\,min
to 240\,min integration time per source to see whether polarized
emission is present before making FR observations.

Only Cyg\,A (Fig.\,1) showed significant linear polarization, having
$1.4\,{\rm mJy\,beam^{-1}}$ polarized emission at the position of the
total intensity peak flux density of $315\,{\rm mJy\,beam^{-1}}$
(0.4~\%). With a detection threshold of 0.3\,\%, all other sources
appeared entirely unpolarized, in the absorbed gaps as well as in all
other locations along their jets.  As the emission process is
undoubtedly synchrotron emission (given the high brightness
temperatures), the lack of polarized emission in these sources needs
explanation.

\begin{figure}
\plotfiddle{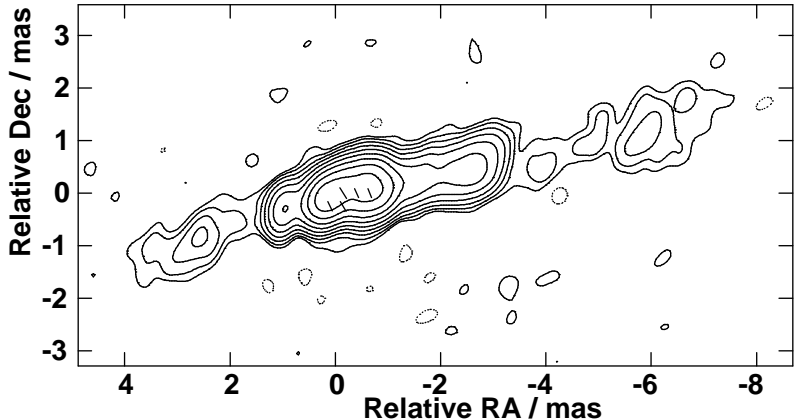}{5cm}{270}{70}{70}{-285}{290}
\caption{Uniformly weighted 15\,GHz image of Cyg\,A with superimposed
polarization vectors.}
\end{figure}

\section{Discussion}

We suggest several intrinsic and extrinsic mechanisms to depolarize
the emission.

{\bf Tangled internal magnetic fields:} Assume the source is optically
thin and its magnetic field is composed of a uniform component $B_0$
and a random component $B_{\rm r}$. Provided that $B_{\rm r}$ varies
on scales much less than the source diameter and that the electrons
have a power-law energy distribution with index $\gamma$, the
intrinsic degree of polarization $p(\gamma)$ will be averaged over the
source to $p_i = p(\gamma)(B^2_0)/(B^2_0 + B^2_r)$ (Burn~1966). For
$\gamma=2$, the intrinsic degree of polarization is 70\,\%. To
depolarize this below our detection threshold, the magnetic field
energy would need to be extremely dominated by the random component,
and so the jets would have to be turbulent, and very little ordering
of the magnetic fields by the overall outward motion of the jet flow
would be permitted. In case the source is optically thick, the maximum
intrinsic degree of polarization is 10\,\% to 12\,\%, and we receive
emission only from the surface. For the source to appear depolarized,
the scale on which the surface magnetic field is tangled must be much
smaller ($<1/10$) than the observing beam. This explanation is
unsatisfactory because the magnetic fields inside the jets must be
ordered to confine them, but the surface must be turbulent, and why
the transition occurs is unexplained.

{\bf Internal Faraday rotation:} In the transition region between
optically thick and optically thin source parts, internal FR could be
significant. Polarized emission from various depths along the line of
sight through the source are Faraday rotated by the source itself, the
degree of rotation depending on the depth of the emitting
region. However, internal FR is also unsatisfactory for explaining our
observed lack of polarization because it requires a significant
fraction of ``cold'' ($\Gamma_{\rm min}\approx1-10$) electrons in the
jet. Internal FR produces a characteristic dependence of polarization
fraction on frequency, but unfortunately, we are not able to test for
the expected wavelength dependence because we lack the required
measurements of the polarization fraction at several wavelengths.

{\bf Bandwidth depolarization:} This depolarization mechanism requires
very high, homogeneous RM ($10^6\,{\rm rad\,m^{-2}}$), and such
conditions should also produce strong FFA which is not seen along most
of the depolarized jets. A possible way out is if the Faraday
screen/absorber is very extended or hot, or both.

{\bf Beam depolarization:} If the magnetic field in a foreground
Faraday screen is tangled on scales much smaller than the observing
beam, regions with similar degrees of polarization but opposite signs
will average out and the observed degree of polarization will be
decreased.  Thus, with spatially highly variable RM, one could in
principle depolarize the source, although the {\it changes} in RM from
region to region still need to be of the order of $10^4\,{\rm
rad\,m^{-2}}$ to depolarize the 15\,GHz band. If one requires at least
10 cells across the beam, the typical cell sizes in NGC\,1052,
NGC\,4261, Cen\,A, Hyd\,A and Cyg\,A need to be 0.01\,pc, 0.02\,pc,
0.002\,pc, 0.12\,pc and 0.13\,pc, respectively. Since these are the
least exotic conditions required by any of the mechanisms discussed,
we feel that beam depolarization in an external medium is the most
likely mechanism to depolarize the sources presented here.

\section{Summary}

Compact core-dominated AGN such as BL Lac objects and quasars
typically display linear polarization of a few to a few tens of
percent on pc scales, with the degree of polarization occasionally
approaching the theoretical maximum of 70\,\%. Thus, some AGN jets
have comparatively modest depolarization. In contrast, the pc-scale
structures of the five radio sources considered here exhibit strong
depolarization at high frequencies. The weakness of their pc-scale
polarization indicates that most likely tangled jet magnetic fields on
sub-pc scales in a foreground screen causes beam depolarization,
although we lack enough information to conclusively decide among the
discussed mechanisms. Why that screen is present in these objects but
absent in most core-dominated AGNs might be an orientation-selection
effect. Core-dominated objects are viewed preferentially pole-on, and
then the lack of depolarization means that the Faraday screen lies in
the equatorial plane. The sample of jets selected here is viewed
preferentially face-on, making the Faraday-screen viewed against the
source.

\end{document}